\documentstyle[psfig,preprint,prl,aps]{revtex}

\newcommand{\be}{\begin{equation}}
\newcommand{\ee}{\end{equation}}
\newcommand{\ba}{\begin{eqnarray}}
\newcommand{\ea}{\end{eqnarray}}

\draft

\begin{document}
\title{Blackbody radiation in a nonextensive scenario}
\author{S. Mart\'{\i}nez${^{1,\,2}}$\thanks{%
E-mail: \tt  martinez@venus.fisica.unlp.edu.ar}, F.
Pennini${^{1,\,2}}$\thanks{ E-mail: \tt 
pennini@venus.fisica.unlp.edu.ar}, A. Plastino${^{1,\,2}}$\thanks{
E-mail: \tt plastino@venus.fisica.unlp.edu.ar}, and C.
Tessone${^1}$\thanks{ E-mail: \tt tessonec@venus.fisica.unlp.edu.ar}.}
\address{$1$ Instituto de F\'{\i}sica de La Plata, National University La Plata, C.C.
727 ,\\ 1900 La Plata, Argentina. \\ $^2$ Argentine
 National Research Council (CONICET)}
\maketitle

\begin{abstract}
An exact analysis of the N-dimensional blackbody radiation process in a
nonextensive \`a la Tsallis scenario is performed for values of the
nonextensive's index in the range ($0<q<1$). The recently advanced ``Optimal
Lagrange Multipliers" (OLM) technique has been employed. The results are
consistent with those of the extensive, $q=1$ case. The generalization of
the celebrated laws of Planck, Stefan-Boltzmann, and Wien are investigated.

\vspace{0.2 cm} PACS: 05.30.-d, 95.35.+d, 05.70.Ce, 75.10.-b

KEYWORDS: Tsallis Thermostatistics, Blackbody radiation.\vspace{1
cm}
\end{abstract}


\newpage

\section{Introduction}
Planck's blackbody radiation studies constitute one of the
milestones in the history of Physics. Planck's law is
satisfactorily accounted by recourse of Bose-Einstein
statistics. The possible existence of small deviations from this
law in the cosmic blackbody radiation has recently been discussed.
They would have arisen, at the time of the matter-radiation
de-coupling \cite{cn1} as a consequence of the fact that
long-range interactions seem to be associated to a nonextensive
scenario \cite {pla2,pla4}. The concomitant thermostatistical
treatment \cite {pla2,pla4,t01,t1,review,t03,t3,pla1,pla3} is by
now recognized as a new paradigm for statistical mechanical
considerations. It revolves around the concept of Tsallis'
information measure $S_{q}$ \cite{t01}, a generalization of
Shannon's one, that depends upon a real index $q$ and becomes
identical to Shannon's measure for the particular value $q=1$.

The blackbody radiation problem was discussed within the Tsallis'
nonextensive framework in \cite{cn1}. It was there conjectured that the
deviations from Planck's law detected in the pertinent data \cite{FIRAS}
could be attributed to a long-range gravitational influence. The study
reported in \cite{cn1} employs the so-called Curado-Tsallis unnormalized
expectation values \cite{t3} in the limit $q\rightarrow 1$. Afterwards,
Lenzi \cite{Lenzi} advanced an exact treatment of the problem, also within
the Curado-Tsallis framework.

Nowadays, it is believed that this framework has been superseded
by the so-called normalized one, advanced in \cite{mendes,pennini},
which seems to exhibit important advantages \cite{review}. This
normalized treatment, in turn, has been considerably improved by
the so-called ``Optimal Lagrange Multipliers'' (OLM) approach
\cite{OLM}. It is then natural to revisit the problem from such
a new context.

In such a vein we shall consider blackbody radiation in
equilibrium within an enclosure of volume $V$ with the goal of
ascertaining the possible $q$-dependence of the Planck spectrum,
Steffan-Boltzmann's law and Wien's one. The OLM treatment seems to
be indicated in view of the findings of \cite{ley0,Rama} regarding
the particular nature of the Lagrange multiplier associated to the
temperature.

The paper is organized as follows: In Sect. \ref{solm} we present
a brief OLM primer, while in Sect. III we obtain the partition
function, internal energy and energy density of our problem. In
Sect. IV first-order corrections in $1-q$ are discussed and,
finally, some conclusions are drawn in Sect. \ref {conclusiones}.

\section{Main results of the OLM formalism}

\label{solm} For a most general quantal treatment, in a
basis-independent way, consideration is required of the density
operator $\hat{\rho}$ that maximizes Tsallis' entropy
\begin{equation}
\frac{S_{q}}{k}=\frac{1-Tr(\hat{\rho}^{q})}{q-1},  \label{entropia}
\end{equation}
subject to the $M$ generalized expectation values $\left\langle \widehat{O}%
_{j}\right\rangle $, where $\widehat{O}_{j}$ $(j=1,\ldots ,M)$ denote the
relevant observables.

Tsallis' normalized probability distribution \cite{mendes}, is obtained by
following the well known MaxEnt route \cite{katz}. Instead of effecting the
variational treatment of Tsallis-Mendes-Plastino (TMP) \cite{mendes}, we
pursue the alternative path developed in \cite{OLM}. One maximizes Tsallis'
generalized entropy (\ref{entropia}) \cite{t01,t1,t2} subject to the
constraints \cite{t01,OLM}

\begin{eqnarray}
Tr(\hat{\rho}) &=&1 \\
Tr\left[ \hat{\rho}^q\left( \widehat{O}_j- \left\langle\widehat{O}%
_j\right\rangle _q\right) \right] &=&0,  \label{vinculos}
\end{eqnarray}
whose generalized expectation values \cite{mendes}

\begin{equation}  \label{gener}
\left\langle\widehat{O}_j \right\rangle _q = \frac{Tr(\hat \rho^q \widehat{O}%
_j)}{Tr(\hat \rho^q)},
\end{equation}
are (assumedly) a priori known. The resulting density operator reads \cite
{OLM}

\begin{equation}
\hat{\rho}=\bar{Z}_q^{-1}\left[ 1-(1-q)\sum_j^M\,\lambda _j\left( \widehat{O}%
_j- \left\langle \widehat{O}_j\right\rangle _q\right) \right] ^{\frac 1{1-q}%
},  \label{rho}
\end{equation}
where $\{\lambda_j\}$ stands for the Optimal Lagrange Multipliers' set and $%
\bar{Z}_q$ is the partition function

\begin{equation}
\bar{Z}_{q}=Tr\left[ 1-(1-q)\sum_{j}^{M}\lambda _{j}\left( \widehat{O}%
_{j}-\left\langle \widehat{O}_{j}\right\rangle _{q}\right) \right] ^{\frac{1%
}{1-q}}.  \label{Zqp}
\end{equation}

It is shown in \cite{OLM} that
\begin{equation}
Tr(\hat{\rho}^q)=\bar{Z}_q^{1-q},  \label{relac1}
\end{equation}
and that Tsallis' entropy can be cast as

\begin{equation}
S_{q}=k\;{\rm \ln }_{q}\bar{Z}_{q},  \label{S2}
\end{equation}
with ${\rm \ln }_{q}\bar{Z}_{q}=(1-\bar{Z}_{q}^{1-q})/(q-1)$. These results
coincide with those of TMP \cite{mendes} in their normalized treatment. If,
following \cite{mendes}, we define now
\begin{equation}
\ln _{q}Z_{q}={\rm \ln }_{q}\bar{Z}_{q}-\bar{Z}_{q}^{1-q}\sum_{j}\lambda
_{j}\ \left\langle \widehat{O}_{j}\right\rangle _{q},  \label{lnqz'}
\end{equation}
we are straightforwardly led to \cite{OLM}
\begin{eqnarray}
\frac{\partial }{\partial \left\langle \widehat{O}_{j}\right\rangle _{q}}%
\left( \frac{S_{q}}{k}\right) &=&\bar{Z}_{q}^{1-q}\lambda _{j}
\label{termo1} \\
\frac{\partial }{\partial \lambda _{j}}\left( \ln _{q}Z_{q}\right) &=&-\bar{Z%
}_{q}^{1-q}\left\langle \widehat{O}_{j}\right\rangle _{q}.  \label{termo2}
\end{eqnarray}

Equations (\ref{termo1}) and (\ref{termo2}) are modified Information Theory
relations like those used to build up an {\em \`a la} Jaynes \cite
{katz}, Statistical Mechanics. The basic Legendre-structure relations can be
recovered in the limit $q \rightarrow 1$.

\section{Blackbody radiation}

\subsection{\label{cn}The standard situation $q=1$}

In order to analyze the blackbody radiation in the
generalized statistical context we first consider
the standard situation, $q=1$. We look for the
equilibrium properties of the blackbody
electromagnetic radiation. The appropriate
thermodynamical variables are the volume $V$ and
the temperature $T$ \cite{Huang,pathria}.

The Hamiltonian of the electromagnetic field, in which there are $%
n_{{\bf k},{\bf \epsilon }}$ photons of momentum ${\bf k}$ and polarization ${\bf %
\epsilon }$, is given by
\begin{equation}
\widehat{{\cal H}}=\sum_{_{k,\epsilon }}\hbar \omega \hat{n}_{k,\epsilon },
\label{H}
\end{equation}
where the frequency is $\omega =c\left| {\bf k}\right| $ ($c$ is
light's speed) and $n_{k,\epsilon }=0,1,2,\ldots $, with no
restrictions on $\left\{ n_{{\bf k},{\bf \epsilon} }\right\}.$ The
partition function reads

\begin{equation}
Z_{1}=Tr\left( e^{-\beta \widehat{{\cal H}}}\right) .  \label{Z1}
\end{equation}

For a macroscopic volume we define a density of states in the $n$%
-dimensional space $g_{n}(\omega )=A_{n}\omega ^{n-1}$ with

\begin{equation}
A_{n}=\frac{2\tau _{n}V}{(4\pi c^{2})^{n/2}\Gamma (n/2)},  \label{An}
\end{equation}
where $\tau _{n}=n-1$ is the number of linear-independent polarizations. Eq.
(\ref{Z1}) can be written as
\begin{equation}
Z_{1}=\exp \left\{ \int_{0}^{\infty }d\omega g_{n}(\omega )\ln \left[ 1-\exp
\left( -\beta \hbar \omega \right) \right] \right\} =e^{\xi _{n}},
\label{Z1f}
\end{equation}
where
\begin{equation}
\xi _{n}=\frac{I_{n}A_{n}}{(\hbar \beta )^{n}},  \label{xin}
\end{equation}
with
\begin{equation}
I_{n}=-\int_{0}^{\infty }dxx^{n-1}\ln \left( 1-e^{-x}\right) =\Gamma
(n)\zeta (n+1),
\end{equation}
$\zeta $ stands for the Riemann z-function and $\Gamma $ is the Gamma
function.

\subsection{Generalized situation $q\neq 1$ ($0<q<1$)}

\subsubsection{Generalized Partition Function}

The OLM-Tsallis generalized partition function (see Section \ref{solm}) is
\begin{equation}
\bar{Z}_{q}=Tr\left[ 1-(1-q)\beta \left( \widehat{H}-U_{q}\right) \right] ^{%
\frac{1}{1-q}},  \label{Zqb}
\end{equation}
where $\widehat{H}$ is the Hamiltonian (\ref{H}), $U_{q}$ is the
mean energy (\ref{gener}), and $\beta $ is the optimal multiplier
Lagrange. The photons' chemical potential is, of course, taken as
zero.

With the aim of calculating $\bar{Z}_{q}$ we follow the steps of
\cite{Lenzi}. We work again in the large volume limit and regard
Eq. (\ref{Zqb})  as an integral that can be evaluated using the
relation \cite{Gradshteyn}

\begin{equation}  \label{pi}
\frac{e^{ab}b^{1-z}}{2\pi }\int\limits_{-\infty }^{\infty }dt\frac{e^{itb}}{%
(a+it)^{z}}=\left\{
\begin{array}{cr}
\frac{1}{\Gamma (z)} & \text{ for }b>0 \\
0 & \text{for }b<0,
\end{array}
\right.
\end{equation}
with $a>0,$ $\mathop{\rm Re}(z)>0$, and $-\pi /2<\arg (a+it)<\pi /2.$

If we set $a=1$, $b=1-(1-q)\beta (\hat{H}-U_{q})$ (the cut-off
condition is naturally fulfilled \cite{review}) and $z=1/(1-q)+1$,
the generalized partition function adopts the appearance

\begin{equation}
\bar{Z}_{q}(\beta )=\int\limits_{-\infty }^{\infty }dtK_{q}(t)e^{\tilde{\beta%
}U_{q}}Z_{1}(\tilde{\beta}),  \label{Zq}
\end{equation}
with $Z_{1}$ given by Eq. (\ref{Z1f}),
\begin{equation}
K_{q}(t)=\frac{\Gamma \left[ (2-q)/(1-q)\right] \exp (1+it)}{2\pi
(1+it)^{(2-q)/(1-q)}},  \label{Kq}
\end{equation}
and
\begin{equation}
\tilde{\beta}=(1+it)(1-q)\beta .  \label{bm}
\end{equation}

In order to evaluate the integral in (\ref{Zq}) we expand the exponential
term and obtain
\begin{equation}
\bar{Z}_{q}(\beta )=\sum\limits_{m=0}^{\infty }\frac{\xi _{n}^{m}}{m!}\frac{%
\Gamma [(2-q)/(1-q)]}{2\pi }\int\limits_{-\infty }^{\infty }dt\frac{%
e^{(1+(1-q)\beta U_{q})(1+it)}}{(1+it)^{\frac{2-q}{1-q}+nm}},
\end{equation}
where $\xi _{n}$ is given by Eq. (\ref{xin}).

Using again (\ref{pi}) with $b=1+(1-q)\beta U_{q},$ $a=1,$ and $%
z=(2-q)/(1-q)+nm$ we arrive to
\begin{equation}
\bar{Z}_q(\beta)=\sum\limits_{m=0}^{\infty }\frac{\xi _{n}^{m}}{m!}\frac{\Gamma [(2-q)/(1-q)]%
}{\Gamma [(2-q)/(1-q)+nm]}\left[ 1+(1-q)\beta U_{q}\right] ^{\frac{1}{1-q}%
+nm},  \label{Zqf}
\end{equation}
and notice that an additional cut-off's like condition must be considered  $%
1+(1-q)\beta U_{q}>0$. Otherwise, $\bar{Z}_{q}(\beta )=0$.

\subsubsection{The internal energy}

The generalized internal energy can be cast in the fashion
\begin{equation}
U_{q}=\frac{T_{1}}{T_{2}},
\end{equation}
where we have introduced the definitions
\begin{eqnarray}
T_{1} &=&Tr\left\{ \left[ 1-(1-q)\beta (\widehat{H}-U_{q})\right] ^{\frac{q}{%
1-q}}\widehat{H}\right\},  \label{A} \\
T_{2} &=&Tr\left\{ \left[ 1-(1-q)\beta (\widehat{H}-U_{q})\right] ^{\frac{q}{%
1-q}}\right\} .  \label{B}
\end{eqnarray}

We first analyze $T_2$. The expression $\left[ 1-(1-q)\beta
(\widehat{H}-U_{q})\right] ^{\frac{q}{1-q}}$ is
evaluated by recourse to (\ref{pi}), where we set $a=1,$ $b=1-(1-q)\beta (%
\widehat{H}-U_{q}),$ and $z=1/(1-q).$ We are thus led to
\begin{equation}
\left[ 1-(1-q)\beta (\widehat{H}-U_{q})\right] ^{\frac{q}{1-q}}=\frac{\Gamma
[1/(1-q)]}{2\pi }\int\limits_{-\infty }^{\infty }dt\frac{e^{(1+it)[1-(1-q)%
\beta (\widehat{H}-U_{q})]}}{(1+it)^{\frac{1}{1-q}}}.
\end{equation}

The trace operation yields afterwards
\begin{equation}
T_{2}=\frac{\Gamma [1/(1-q)]}{2\pi }\int\limits_{-\infty }^{\infty }dt\frac{%
e^{(1+it)[1+(1-q)\beta U_{q}]}}{(1+it)^{\frac{1}{1-q}}}Z_{1}(\tilde{\beta}),
\end{equation}
with $\tilde{\beta}$ defined by Eq. (\ref{bm}).  Expanding exponents and
employing once again (\ref{pi}) leads finally to
\begin{equation}
T_{2}=\sum\limits_{m=0}^{\infty }\frac{\xi _{n}^{m}}{m!}\frac{\Gamma
[1/(1-q)]}{\Gamma [1/(1-q)+nm]}\left[ 1+(1-q)\beta U_{q}\right] ^{\frac{q}{%
1-q}+nm}.
\end{equation}

As for $T_2$, we recast $T_1$ as
\begin{equation}
T_1=\frac{\Gamma [1/(1-q)]}{2\pi }\int\limits_{-\infty }^{\infty }dt\frac{%
e^{(1+it)[1+(1-q)\beta U_{q}]}}{(1+it)^{\frac{1}{1-q}}}Tr\left[ Z_{1}(\tilde{%
\beta})\widehat{H}\right].
\end{equation}
We proceed as follows: i) take advantage of the
fact that
\begin{equation}
Tr\left[ Z_{1}(\tilde{\beta})\widehat{H}\right] =-\frac{\partial Z_{1}(%
\tilde{\beta})}{\partial \tilde{\beta}},  \label{traza}
\end{equation}
 ii) use Eq. (\ref{Z1f}), and iii)  expand again
 the exponential
 term.  We find
\begin{equation}
T_1=\frac{n \Gamma [1/(1-q)]}{2\pi (1-q)\beta }\sum\limits_{m=0}^{\infty }%
\frac{\xi _{n}^{m+1}}{m!}\int\limits_{-\infty }^{\infty }dt\frac{%
e^{(1+it)[1+(1-q)\beta U_{q}]}}{(1+it)^{\frac{1}{1-q}+n(m+1)}}.
\end{equation}

\begin{figure}[tbp]
\psfig{file=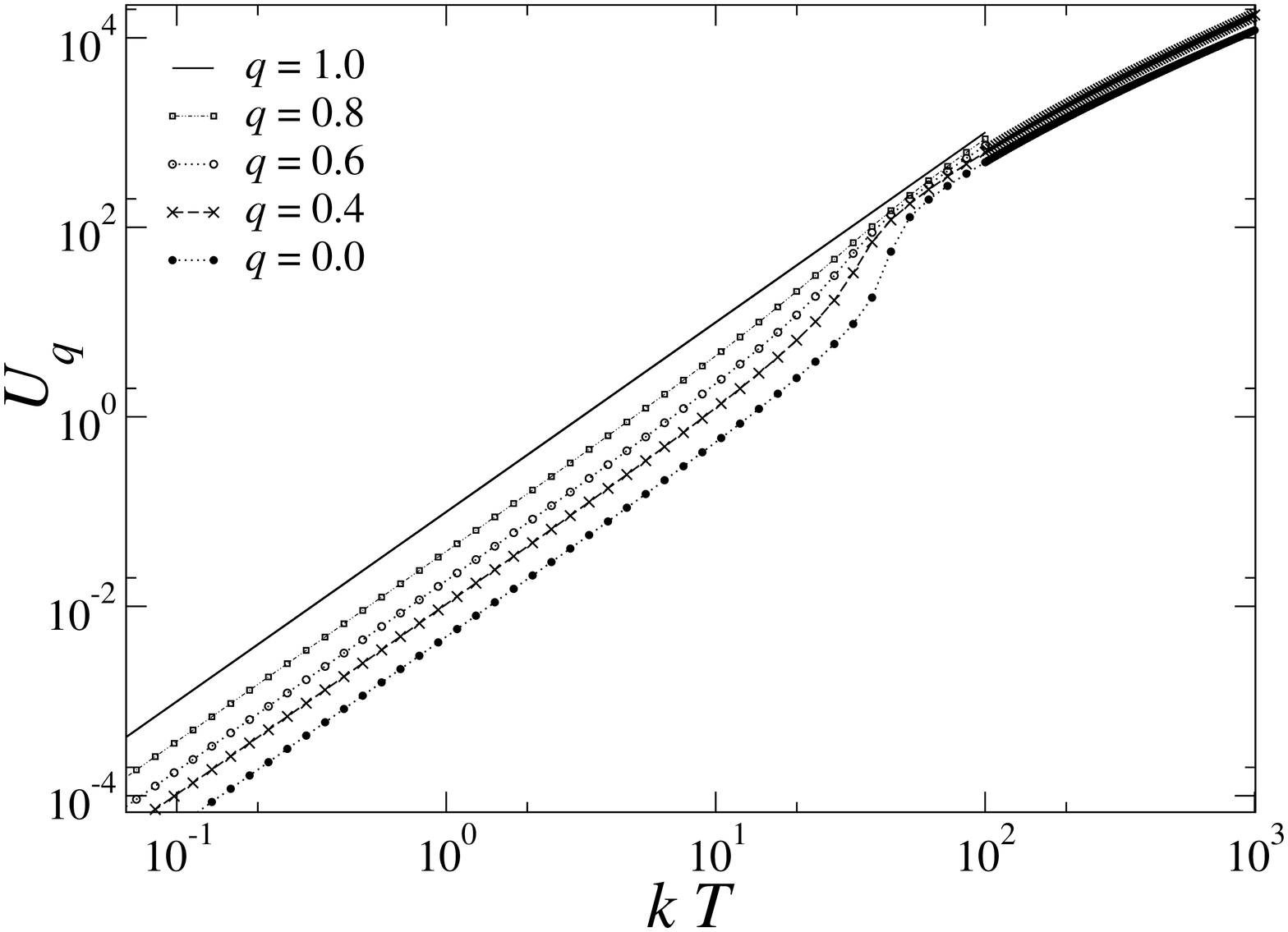,width=10cm,angle=-90,silent=true}
\caption{The internal energy $U_q$ as a function of $T=1/\beta$ for
different values of the non-extensivity parameter $q$.}
\label{Uq-T}
\end{figure}

The integral is evaluated as before, which yields
\begin{equation}
T_{1}=\frac{n}{(1-q)\beta }\sum\limits_{m=0}^{\infty }\frac{\xi _{n}^{m+1}}{%
m!}\frac{\Gamma [1/(1-q)]}{\Gamma [1/(1-q)+n(m+1)+1]}\left[ 1+(1-q)\beta
U_{q}\right] ^{\frac{1}{1-q}+n(m+1)},
\end{equation}
so that the internal energy becomes
\begin{equation}
U_{q}=\frac{n\xi _{n}}{(1-q)\beta }\frac{\sum\limits_{m=0}^{\infty }\frac{%
\xi _{n}^{m}}{m!}\frac{1}{\Gamma [1/(1-q)+n(m+1)+1]}\left[ 1+(1-q)\beta
U_{q}\right] ^{n(m+1)+1}}{\sum\limits_{m=0}^{\infty }\frac{\xi _{n}^{m}}{m!}%
\frac{1}{\Gamma [1/(1-q)+nm]}\left[ 1+(1-q)\beta U_{q}\right] ^{nm}},
\label{Uq}
\end{equation}
although compliance with the Tsallis'cut-off condition $1+(1-q)\beta U_{q}>0$
is always to be demanded. The above is a non-linear equation to be tackled
in numerical fashion. Some pertinent results are displayed in Fig. \ref{Uq-T}, for different $q$-values, they include the standard $q=1$ case
that yields the {\it Steffan Boltzmann law}
\begin{equation}
U_{1}=\sigma T^{4}
\end{equation}
with $\sigma =5.67\times 10^{-8}W/m^{2}\ K^{4}$, the so-called {\it 
 Steffan-Boltzmann constant}.

\begin{figure}[tbp]
\psfig{file=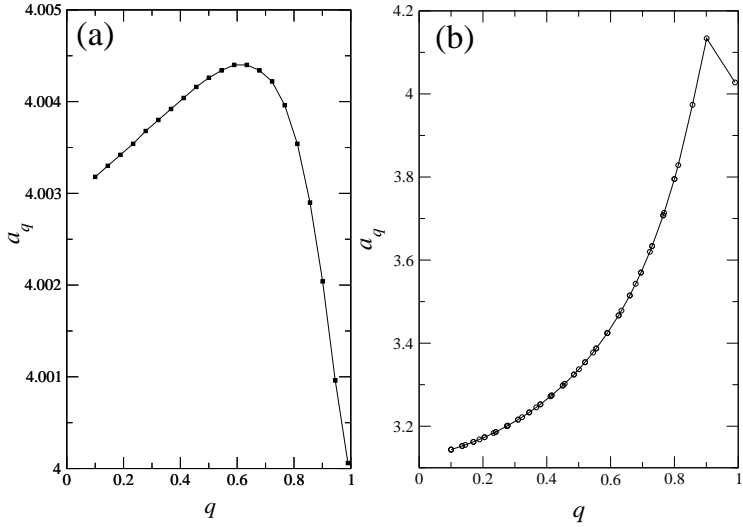,width=10cm,angle=-90,silent=true}
\caption{Power law exponent $a_q$ as a function of $q$ (see text
for details). The graph (a) corresponds to small temperatures, and
the (b)-one to high $T$-values.} \label{aq-q}
\end{figure}

We notice that a Steffan Boltzmann-like law holds, if $kT$ is small enough,
for all $q \in (0<q<1)$, with $\sigma=\sigma(q)$. For high $T$-values and $q\ll 1$ the power law behavior seems to persist, but the corresponding power, let us call it $a_{q}$, is no longer equal to $4$. Fig. \ref{aq-q} is a graph of $a_{q}$ versus $q$ for
\begin{equation}
U_{q}\propto T^{a_{q}}.
\end{equation}

Fig. \ref{aq-q} (a) shows $a_{q}$ for small $kT$-values. The
validity of Stefan Boltzmann's Law for a wide range of $q$-values
is easily appreciated. Fig. \ref{aq-q} (b) depicts, instead,
$a_{q}$ for large $kT$-values. In this limit, important deviations
from the $a_{q}=4$ value are detected.

An intermediate temperature range is observed in Fig. \ref{Uq-T}
in which the power law behavior (of $U_{q}$ vs. $kT$) is violated.
The corresponding transition is the more abrupt the larger $\vert
q-1 \vert$. As will be seen in the next section, these
violations can be attributed to a Tsallis cut-off in the energy
densities.

\subsubsection{Energy densities}

The generalized spectral energy distribution of blackbody radiation $u_{q}$
is given by
\begin{equation}
U_{q}=\int_{0}^{\infty }d\omega u_{q}.  \label{uq}
\end{equation}

In order to obtain $u_q$ we work with $U_q=T_1/T_2$, with $T_1$ and $T_2$
given by Eqs. (\ref{A}) and (\ref{B}), respectively. The denominator will
not suffer any change, but in the numerator we will keep the integral form
of $Z_1$ as given in Eq. (\ref{Z1f}). Then, Eq. (\ref{traza}) will lead to
\begin{equation}
Tr \left( e^{-\tilde \beta \widehat{H}} \widehat{H}\right)= \hbar
A_{n}e^{\xi_n}\int\limits_{0}^{\infty }d\omega \omega ^{n}\frac{e^{-\tilde{%
\beta}\hbar \omega }}{1-e^{-\tilde{\beta}\hbar \omega }},
\end{equation}
allowing us to cast $T_1$ in the fashion
\begin{equation}
T_{1}=\frac{\Gamma [1/(1-q)]}{2\pi }\hbar A_{n}\int\limits_{0}^{\infty
}d\omega \omega ^{n}\int\limits_{-\infty }^{\infty }dt\frac{%
e^{(1+it)[1-(1-q)\beta (\hbar \omega -U_{q})]}}{(1+it)^{\frac{1}{1-q}}}\frac{%
e^{\xi _{n}}}{1-e^{-\tilde{\beta}\hbar \omega }}.
\end{equation}

Recourse to the identity
\begin{equation}
\frac{1}{1-e^{-\tilde{\beta}\hbar \omega }}=\sum\limits_{s=0}^{\infty
}\left( e^{-\tilde{\beta}\hbar \omega }\right) ^{s},
\end{equation}
yields now
\begin{equation}
T_{1}=\frac{\Gamma [1/(1-q)]}{2\pi }\hbar A_{n}\int\limits_{0}^{\infty
}d\omega \omega ^{n}\sum\limits_{s=0}^{\infty }\sum\limits_{m=0}^{\infty }%
\frac{\xi _{n}^{m}}{m!}\int\limits_{-\infty }^{\infty }dt\frac{%
e^{(1+it)[1-(1-q)\beta [\hbar \omega (1+s)-U_{q}]]}}{(1+it)^{\frac{1}{1-q}%
+nm}}.
\end{equation}

If we use Eq. (\ref{pi}) the integral above is easily calculated, and we
obtain
\begin{equation}
T_{1}=\Gamma [1/(1-q)]\hbar A_{n}\int\limits_{0}^{\infty }d\omega \omega
^{n}\sum\limits_{s=0}^{\infty }\sum\limits_{m=0}^{\infty }\frac{\xi _{n}^{m}%
}{m!}\frac{\left[ 1-(1-q)\beta [\hbar \omega (1+s)-U_{q}]\right] ^{\frac{q}{%
1-q}+nm}}{\Gamma [1/(1-q)+nm]}.
\end{equation}

According to (\ref{uq}) the ratio $\frac{T_1}{T_2}$ implies that
\begin{equation}
u_{q}=\hbar A_{n}\omega ^{n}\frac{\sum\limits_{m=0}^{\infty }B_m S_m}{%
\sum\limits_{m=0}^{\infty }B_m},  \label{uqf}
\end{equation}
where
\begin{eqnarray}
B_m&=&\frac{\xi _{n}^{m}}{m!}\frac{\left[ 1+(1-q)\beta U_{q}\right] ^{nm}}{%
\Gamma [1/(1-q)+nm]}, \\
S_m&=&\sum\limits_{s=1}^{\infty }\left[ 1- \frac{(1-q)\beta [\hbar \omega
(s+1)-U_{q}]}{1+(1-q)\beta U_q}\right] ^{\frac{q}{1-q}+nm}.  \label{Sm}
\end{eqnarray}

\begin{figure}[tbp]
\psfig{file=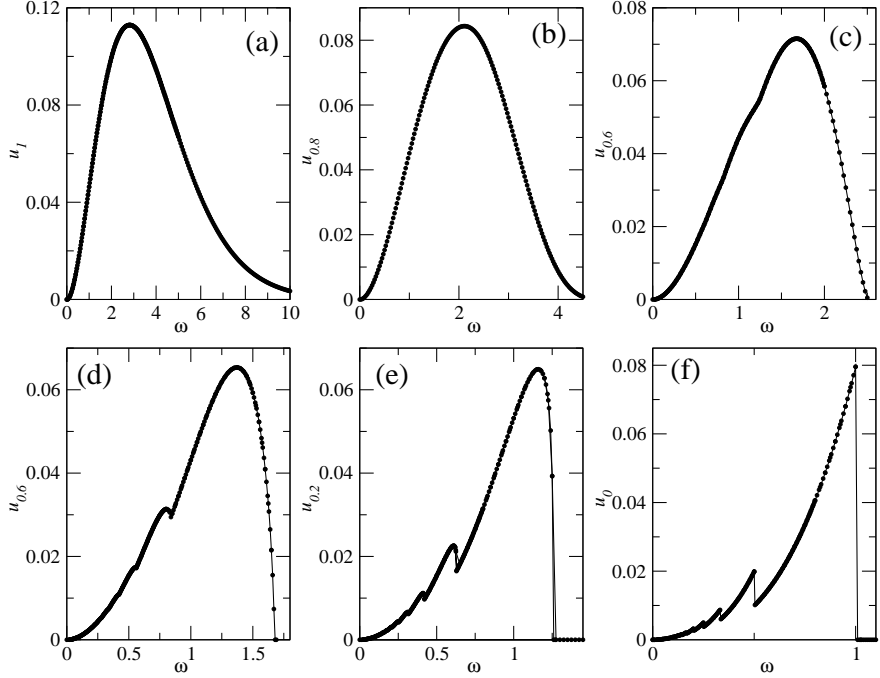,width=12cm,angle=-90,silent=true}
\caption{Energy density $u_q$ as a function of the frequency $\omega $. Each
plot corresponds to a different value for $q$. $k T$ equals unity in all
cases}
\label{dens-T1}
\end{figure}

\begin{figure}[tbp]
\psfig{file=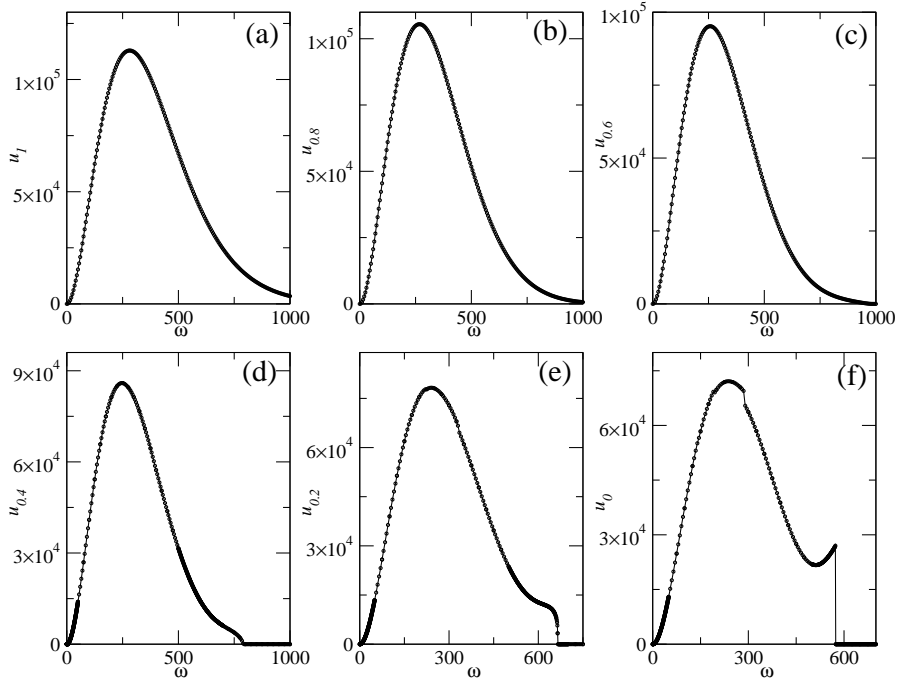,width=12cm,angle=-90,silent=true}
\caption{Energy density $u_q$ as a function of the frequency $\omega $. Each
plot correspond to different values of $q$ parameter. In all of them, $k T
=10^2$ was kept fixed.}
\label{dens-T100}
\end{figure}

\begin{figure}[tbp]
\psfig{file=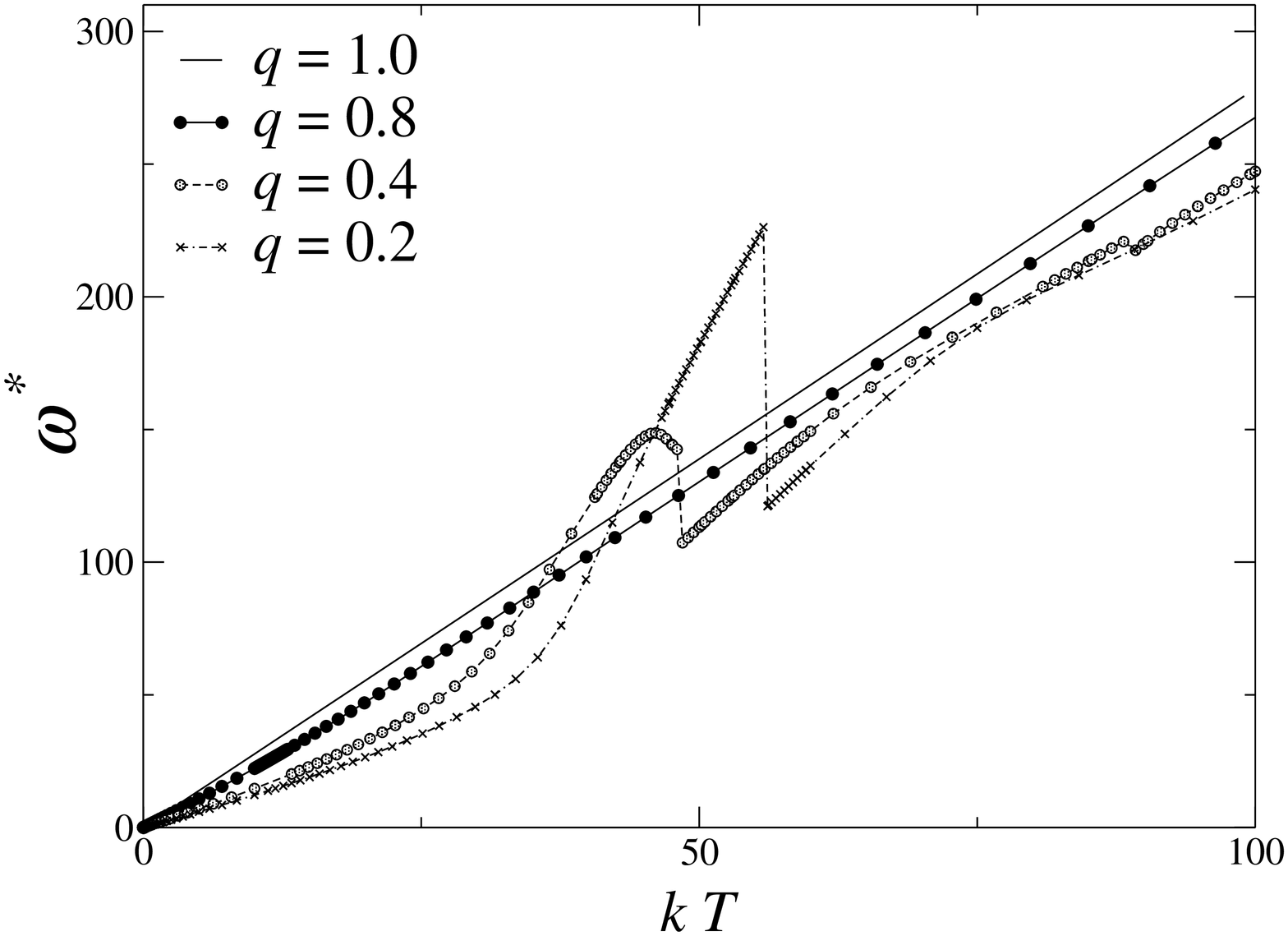,width=12cm,angle=-90,silent=true}
\caption{We depict the frequency $\omega^*$ for which the energy density reaches its
maximum, as a function of $k T$, for different
values of $q$.}
\label{fWein}
\end{figure}

Figs. \ref{dens-T1} and \ref{dens-T100} are plots of $u_{q}$ vs. the
frequency $\omega $. Notice that they quite resemble the semi-empirical
Planck law for small $\vert q-1 \vert $-values.  Fig. \ref{fWein} clearly
shows that the frequency shifts of the maxima  of the above referred to
curves (located at $\omega=\omega^*$) comply with the {\it Wien law}
\begin{equation}
\omega ^{*}\propto T,
\end{equation}
for such $q$-values. Up to $q=0.8$ this nice behaviour persists.
As $q$ departs from $1$ an anomalous behaviour ensues. Local
maxima begin to appear in the curves $u_{q}(\omega )$ (Figs.
\ref{dens-T1} and \ref{dens-T100}). As $T$ grows the global
maximum is shifted and the secondary ones become more pronounced,
until one of them replaces the former one as the maximum
maximorum. An abrupt jump in the curve $\omega ^{*}$ vs. $T$
ensues as a consequence (Fig. \ref{fWein}).

The departures from Planck's law as $\vert q-1 \vert$ grows, could
in principle be subject to experimental verification, thus $q$ would
be fixed.

The above referred to discontinuities in the ``Wien-like" behaviour are due
to the Tsallis' cut-off operative in terms of the sums $T_{1}$ and $%
T_{2}$.

\section{First order corrections}

We have seen in the preceding section that (\ref{uqf}) yields, for $q$%
-values close to unity, results that quite resemble the ones given by
Planck's law. It is then reasonable to look for a perturbative expansion in $%
1-q$.

Let us start with $S_m$ (Cf. Eq. (\ref{Sm})) and write it in the form
\begin{equation}
S_m=\sum_{s=0}^{\infty}\exp\left\{ \left( \frac{q}{q-1}+n
m\right)\ln\left[1-(1-q) \frac{\beta \hbar \omega
(1+s)}{1+(1-q)\beta U_q} \right]\right\}.
\end{equation}

A first order expansion of the logarithm yields
\begin{equation}
S_m\approx\sum_{s=0}^{\infty}\exp\left\{ -[ q+n m(1-q)] \frac{\beta \hbar
\omega (1+s)}{1+(1-q)\beta U_q} \right\},
\end{equation}
a power series in $s$ of guaranteed convergence for $1-q<1$.
Neglecting the term $n m(1-q)$ on account of the fact that the
concomitant series rapidly converges so that terms for which $m$
could compete with $1/(1-q)$ are indeed negligible, we have
\begin{equation}
S_m\approx \frac{1}{\exp\left[ \frac{q\beta \hbar \omega }{1+(1-q)\beta U_q}
\right]-1},  \label{Smap}
\end{equation}
and replacement in Eq. (\ref{uqf}) yields

\begin{equation}
u_q\approx \frac{\hbar A_n \omega^n}{\exp\left[ \frac{q\beta \hbar \omega }{%
1+(1-q)\beta U_q} \right]-1},  \label{uqapp}
\end{equation}
a first order correction for a ``generalized
Planck law" that yields the classical result for
$q\rightarrow 1$. Eq. (\ref{uqapp}) provides one
then with an approximate energy density.

Using now this equation as a starting point we also get first order
corrections in $1-q$ to the Wien law. We can locate the maximum with respect to $%
\omega$ of Eq. (\ref{uqapp}) with the auxiliary definition
\begin{equation}
x=\frac{\omega q \beta \hbar}{1-(1-q) \beta U_q},  \label{x}
\end{equation}
and immediately find

\begin{equation}
e^{-x}+\frac{x}{n}=1,
\end{equation}
whose solution is a constant $b$ for each fixed
value of $n$. For $n=3$ we find, for instance,
$b=2.82$. The maximum of $u_{q}(\omega )$ for
different $T$'s is located at distinct frequencies
according to

\begin{equation}
\omega _{i}=\frac{bk}{q\hbar }T_{i}-\frac{1-q}{q}\frac{U_{q}}{\hbar },
\end{equation}
a first order correction to Wien's law.

If we integrate Eq. (\ref{uqapp}) over frequencies we find $U_q$ as a
function of $T$, and a concomitant first order correction to Planck's law

\begin{equation}
U_{q}=\hbar A_{n}\alpha \left[ \frac{1+(1-q)\beta U_{q}}{q\beta \hbar }%
\right] ^{n+1},
\end{equation}
that reduces trivially to Planck's one  in the limit $%
q\rightarrow 1$. We call $\alpha $ the integral

\begin{equation}
\alpha=\int_0^\infty dx \frac{x^n}{e^x-1}=\Gamma(n+1)\zeta(n+1),
\end{equation}
with $x$ given by (\ref{x}).

It may be convenient to point out that the first
order relationships here obtained ignore possible
cut-off problems. Anyway, in the $0<q<1$ range
 (with $q \rightarrow 1$) that is relevant to the
present considerations, such problems should not
arise.

\section{Conclusions}

\label{conclusiones}We have presented here exact results for the
mean energy and the energy density of the blackbody problem in a
nonextensive environment. A {\it Stefan Boltzmann-like law } is obtained,
for all $0<q<1$%
, although with a $q$-dependent SB constant $\sigma (q)$, for almost the
whole temperature range, although for some temperatures a deviation from the
power law behaviour is detected.

The energy density curves correspond to Planck's law for $q$ values close to
unity. For other $q$'s a gradual departure from the typical Planck curve is
observed, which can be attributed to Tsallis' cut-off effects. {\it Wien's law%
} is also obtained for $q$ values close to unity.

Finally, we have presented first order perturbative results in the limit $%
q\rightarrow 1$ for the energy density $u_{q}$ (the parameter is, of course,
$1-q$). These could be of some utility in obtaining first order nonextensive
corrections to the Bose-Einstein equation.

\smallskip \acknowledgements The financial support of the National Research
Council (CONICET) of Argentina is gratefully acknowledged. F. Pennini
acknowledges financial support from UNLP, Argentina.

%
%

\begin{references}
\bibitem{cn1}  C. Tsallis, F. C. S\`{a} Barreto, and E. D. Loh {\it Physical
Review E} {\bf 52} 1447.

\bibitem{pla2}  A. R. Plastino and A. Plastino, {\it Braz. J. of Phys.} {\bf %
29} (1999) 79.

\bibitem{pla4}  A. R. Plastino, A. Plastino, {\it Physics Letters A} {\bf 174%
} (1993) 834.

\bibitem{t01}  C. Tsallis, {\it Braz. J. of Phys.} {\bf 29} (1999) 1, and
references therein. See also
http://www.sbf.if.usp.br\-/WWW\_pages/Journals/BJP/Vol129/Num1/index.htm

\bibitem{t1}  C. Tsallis, {\it Chaos, Solitons, and Fractals} {\bf 6} (1995)
539, and references therein; an updated bibliography can be found in
http://tsallis.cat.cbpf.br/biblio.htm

\bibitem{review}  C. Tsallis, Nonextensive statistical mechanics and
thermodynamics: Historical background and present status, in ``Nonextensive
Statistical Mechanics and its Applications'', eds. S. Abe and Y. Okamoto,
``Lecture Notes in Physics'' (Springer-Verlag, Berlin, 2000), in press.

\bibitem{t03}  C. Tsallis, {\it Physics World 10} (July 1997) 42.

\bibitem{t3}  E.M.F. Curado and C. Tsallis, {\it J. Phys. A} {\bf 24}
(1991) L69; Corrigenda: {\bf 24} (1991) 3187 and {\bf 25} (1992)
1019.


\bibitem{pla1}  A. Plastino and A. R. Plastino, {\it Braz. J. of Phys.} {\bf
29} (1999) 50.


\bibitem{pla3}  A. R. Plastino and A. Plastino, {\it Phys. Lett. A} {\bf 177}
(1993) 177.


\bibitem{FIRAS}  J. C. Mather et {\it al.}, {\it Astrophys. J}. {\bf 420}
(1994) 439

\bibitem{Lenzi}  E. K. Lenzi, R. S. Mendes, {\it Physics Letters A} {\bf 250
} (1998) 270.

\bibitem{mendes}  C. Tsallis, R. S. Mendes, and A. R. Plastino, {\it Physica
A} {\bf 261} (1998) 534.

\bibitem{pennini}  F. Pennini, A. R. Plastino and A. Plastino, {\it Physica
A } {\bf 258} (1998) 446.

\bibitem{OLM}  S. Mart\'{\i }nez, F. Nicol\'{a}s, F. Pennini and A.
Plastino, {\it Physica A} (2000) (en prensa).

\bibitem{ley0}  S. Mart\'{\i }nez, F. Pennini and A. Plastino, preprint
(2000) [cond-math/0004448].

\bibitem{Rama}  S. K. Rama, preprint 2000 [cond-mat/0006279].

\bibitem{katz}  E. T. Jaynes in {\it Statistical Physics}, ed. W. K. Ford
(Benjamin, NY, 1963); A. Katz, {\it Statistical Mechanics},
(Freeman, San Francisco, 1967).

\bibitem{t2}  C. Tsallis, {\it J. Stat. Phys.} {\bf 52} (1988) 479.

\bibitem{Huang}  K. Huang, Statistical Mechanics (Wiley, New York, 1987) pp.
278-283.

\bibitem{pathria}  R. K. Pathria, Statistical Mechanics, Pergamon, New York,
1985.

\bibitem{Gradshteyn}  I. S. Gradshteyn, I. M. Ryzhik, Table of Integrals
Series and Products (Academic Press, New York, 1980) p. 935.
\end{references}
\end{document}